\documentclass[onecolumn,nopacs,floatfix,superscriptaddress]{revtex4}
\pdfoutput=1 
\usepackage[utf8]{inputenc}
\usepackage{hyperref}
\usepackage{amsmath}
\usepackage{graphicx}  
\usepackage{color}

\begin{document}

\title{Beehive scale-free emergent dynamics}

\author{Ivan Shpurov,\textsuperscript{1}\footnote[1]{Electronic address: ivan.shpurov@oist.jp} Tom Froese}   
\affiliation{Okinawa Institute of Science and Technology Graduate University, Embodied Cognitive Science Unit, Tancha, Okinawa, Japan}
\author{Dante R. Chialvo}
\affiliation{Instituto de Ciencias F\'isicas (ICIFI-CONICET), Center for Complex Systems and Brain Sciences (CEMSC3), Escuela de Ciencia y Tecnolog\'ia, Universidad Nacional de Gral. San Mart\'in, Campus Miguelete, San Mart\'in, Buenos Aires, Argentina}
\affiliation{Consejo Nacional de Investigaciones Cient\'{\i}fcas y Tecnol\'ogicas (CONICET), Buenos Aires, Argentina}

\date{November 28, 2023}

\begin{abstract}
 It has been repeatedly reported that the collective dynamics of social insects exhibit universal emergent properties similar to other complex systems.  In this note, we study a previously published  data set in which the positions of thousands of honeybees in a hive are individually tracked over multiple days. The results show that the hive dynamics exhibit long-range spatial and temporal correlations in the occupancy density fluctuations, despite the characteristic short-range bees' mutual interactions.  The variations in the occupancy unveil a non-monotonic function between density and bees' flow, reminiscent of the car traffic dynamic near a jamming transition at which the system performance is optimized  to achieve the highest possible throughput.  Overall, these results suggest that the beehive collective dynamics are self-adjusted towards a point near its optimal density. 
\end{abstract}

\maketitle

 The striking collective phenomena exhibited by social insects and animals have inspired complex systems scientists for decades.  The study of the basic principles of their communication and self-organization into large ensembles can be traced back to the seminal work of Wilson on social insects \cite{Wilson1,Wilson2,Wilson3}.  
Large collective behavioral structures, such as hives, swarms, bird flocks, foraging ants trails, etc.,  emerge out of local interactions. The sizes of the resulting complex global structures are several orders of magnitude larger than the scale that the individuals are able to communicate. Prototypical examples are the large trails formed by foragers ants \cite{rauch}, the collective defensive reactions of entire bird flocks \cite{Cavagna1, Cavagna2} or the shimmering phenomena \cite{shimmering} seen in giant honeybees as waves rippling across their open nests, which has an adaptive value, facilitating the species survival. Often, the dynamics of these apparently  disparate large-scale phenomena exhibit scale-invariant properties both in space and time which are  common across the species, an observation that could be studied  from the perspective of statistical physics \cite{Cavagna3,peleg}.
    
In this work we study the collective dynamics of a hive  of thousands of bees, searching for clues of their large-scale dynamics, revisiting the data previously acquired by Gernat el al. \cite{Gernart} graciously shared with the authors of the current work. The paper is organized as follows: In the next paragraphs,  we describe briefly the main features of the data and the methods. Next, we study the univariate aspects of the fluctuations, including occupancy (density) fluctuations, nearest neighbors bees' interaction distances, and insects' speed. After that, we proceed to explore collective effects, by calculating spatial and temporal correlations for different coarse grainings of the data being considered.
Finally, we investigate, in analogy to traffic dynamics, the relation between average mobility and local density. The paper concludes with a discussion on  the possible origin of the scaling found in the hive collective dynamics. 
\begin{figure}[hb!]
\includegraphics[width=.9\columnwidth]{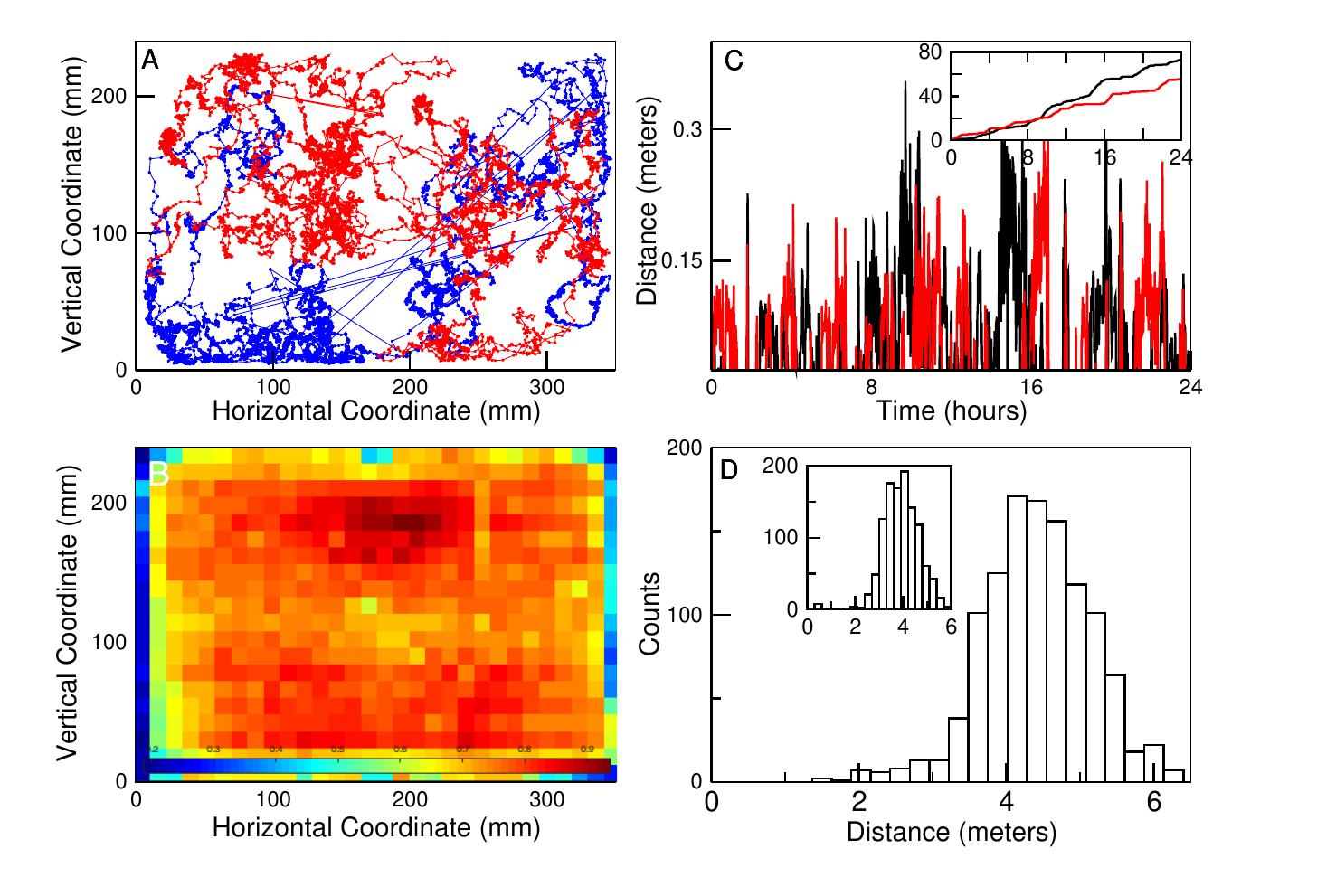}
\centering
 \caption{Typical bees displacements in the hive. Panel A illustrates the trajectory of two bees during the 7 hours of the first day of the experiment. Straight lines represent the approximate position of the bee when it walked on the glass and its exact coordinates are unavailable. Panel B depicts, as a heat map, the average occupancy  over the entire experiment of the hive (discretized in $30\times20$ squares of side length = 11.6 mm). Panel C  shows two examples of the distances covered in each  60-second intervals by two bees on the first day of the experiment. In the main panel, each point represents the distance covered while the inset shows the cumulative distance as a function of time. Panel D illustrates the distribution for all bees of the average distance traveled in one hour computed for the entire duration of the experiment.  The results in the main panel correspond to the statistics computed for the daytime, and those in the inset for the nighttime.} 
\label{fig:Fig1}
\end{figure}

\emph{Data:} The full details of the methods are described in Ref.\cite{Gernart}. In brief, the experiment is conducted with a colony of 1200 one-day-old worker bees with a queen which are placed in a rectangular hive (dimensions $348 \times232$ mm) covered by a transparent material.  The hive thickness is such that prevents the bees from climbing on top of each other while permitting full observation for recording.  The dataset includes the position (the center of mass) of each bee sampled every second. It should be noted that bees could exhibit two types of behaviors that allow them to elude observation and also exclude them from interactions with the rest of the insects in the hive: they could go inside the honeycomb cells to sleep, eat, and take care of the brood or they could walk on the glass cover of the hive. 
For the first two days and two nights of the recording, insects were kept sealed and were supplied with the necessary nutrients (hereafter termed ``phase I''). Afterward, bees were allowed to leave at  will to scout and forage (``phase II''). During the last days of the trial, forages were not permitted to return to the hive (``phase III''). \footnote{The top 16 rows of the honeycomb cells were provisioned with 150 g of honey, and the two rows below the honey cells were provisioned with 15 g of artificial ``bee bread". Each  bee was marked with barcodes attached to their thoraxes, enabling the continuous tracking of their positions. The sampling rate of the raw data is 1 sec.  Five different colonies were recorded, each for several days and nights. Results presented in the  paper are computed using data from the second trial (dataset \#2).}

\emph{Density fluctuations:} Figure \ref{fig:Fig1}A shows an example of the trajectories of two bees during the 7 hours of the first day of the experiment, from which it can be seen that their activity covers the entire hive. This is best appreciated by computing the average density of bees discretizing  the hive in $30\times20$ squares of side length = 11.6 mm and counting the number of bees in each grid cell at a sampling rate of 5 seconds. Figure \ref{fig:Fig1}B depicts, as a heat map, the average occupancy over the entire experiment.
Typically, each bee, unless it is hiding inside one of the honeycomb cells, explores an average distance of  $\sim 4-5$ meters  every hour (see Panel D) displaying bursts of activity  (as described before \cite{Gernart}) seen in the two examples presented in Panel C.

\begin{figure}[hb!]
\includegraphics[width=.95\columnwidth]{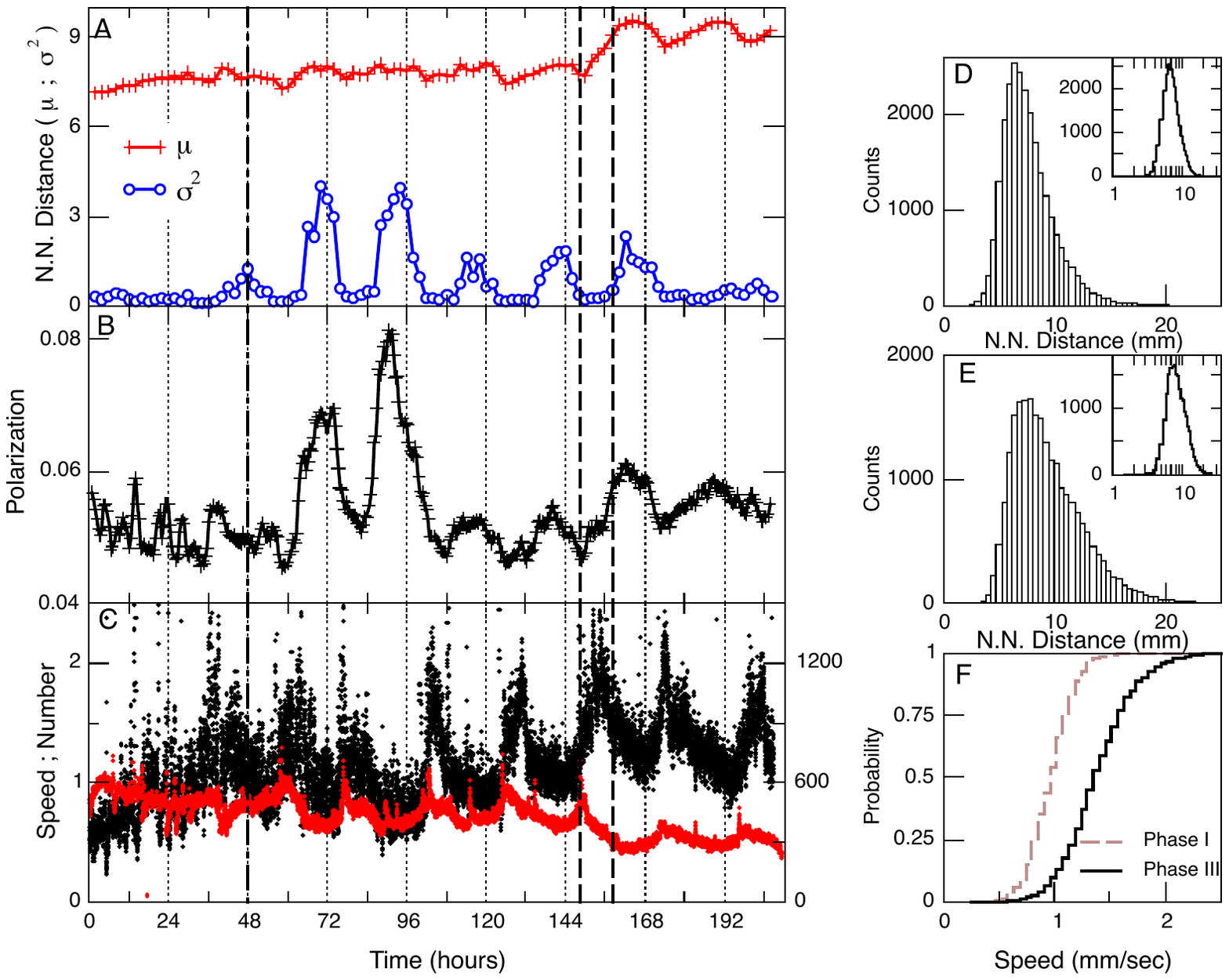}
 \centering
 \caption{Bees nearest neighbor distances over time. Panel A shows the mean $\mu$ and variance $\sigma^2$, computed every two hours for \emph{all} bees for the entire duration of the experiment. The vertical dashed-dotted line at 48 hours indicates the sunrise of the third day when the hive entrance was open and bees were able to go out. The vertical black dashed lines at 149 and 158 hours indicate the moments at which foragers removal has begun/ended. Notice that the peaks in both the mean distance and its variance coincide approximately with each day's sunrise. Panel B shows the  bees motion polarization $\Phi$ (i.e., Eq. \eqref{eq:Polar}), computed every 5 second and plotted here as averages of 2 hours. 
 Panel C shows the mean speed of all bees in the hive computed every 5 seconds throughout the entire experiment. In addition, the  number of bees detected in the field are plotted (with red symbols).  Panels D and E: Histogram of the nearest neighbor distances for a single bee, in linear axis in the main plots and semilogarithmic axis in the insets. The results on panel D correspond to the first 48 hours of data and those in panel E  to 48 hours after the foragers' removal for \emph{one} of the bees that remained in the hive. The cumulative distribution of speeds for \emph{all} the bees detected are depicted in panel F, both for the period preceding the hive opening (i.e., dashed vertical line at 48 hours in main panels, ``phase I'') and for the period of time after the foragers removal ended (i.e., after the hour 158, ``phase III'')}.
\label{fig:Fig2}
\end{figure}

It is well known that bees in the hive usually are highly packed. To study the typical inter-bees interaction length over time we computed (for each bee) its distance to their nearest neighbor (NN) (between their centers of masses) at intervals of 5 seconds. It can be noted that the mean distance  is  of the same order as the typical bee's body length $\mu=10$ mm +/- 2) and that exhibits daily fluctuations as shown in Figure \ref{fig:Fig2}A and B.  The distribution of nearest neighbor distances that a single bee experiences also fluctuates, as illustrated by the histograms shown in Panels D and E of Figure \ref{fig:Fig2}, which seem to be approximated by a log-normal function (see insets showing the histograms plotted in log-linear axis).
As expected the bees' displacement speed also exhibits changes. Notice in panel C the circadian fluctuations in speed as the experiment evolves, as well as the one-fold change at the end of the experiment,  when the bee density in the hive decreases after the foragers were not allowed to return to the hive.

\begin{figure}[]
    \includegraphics[width=.6 \columnwidth]{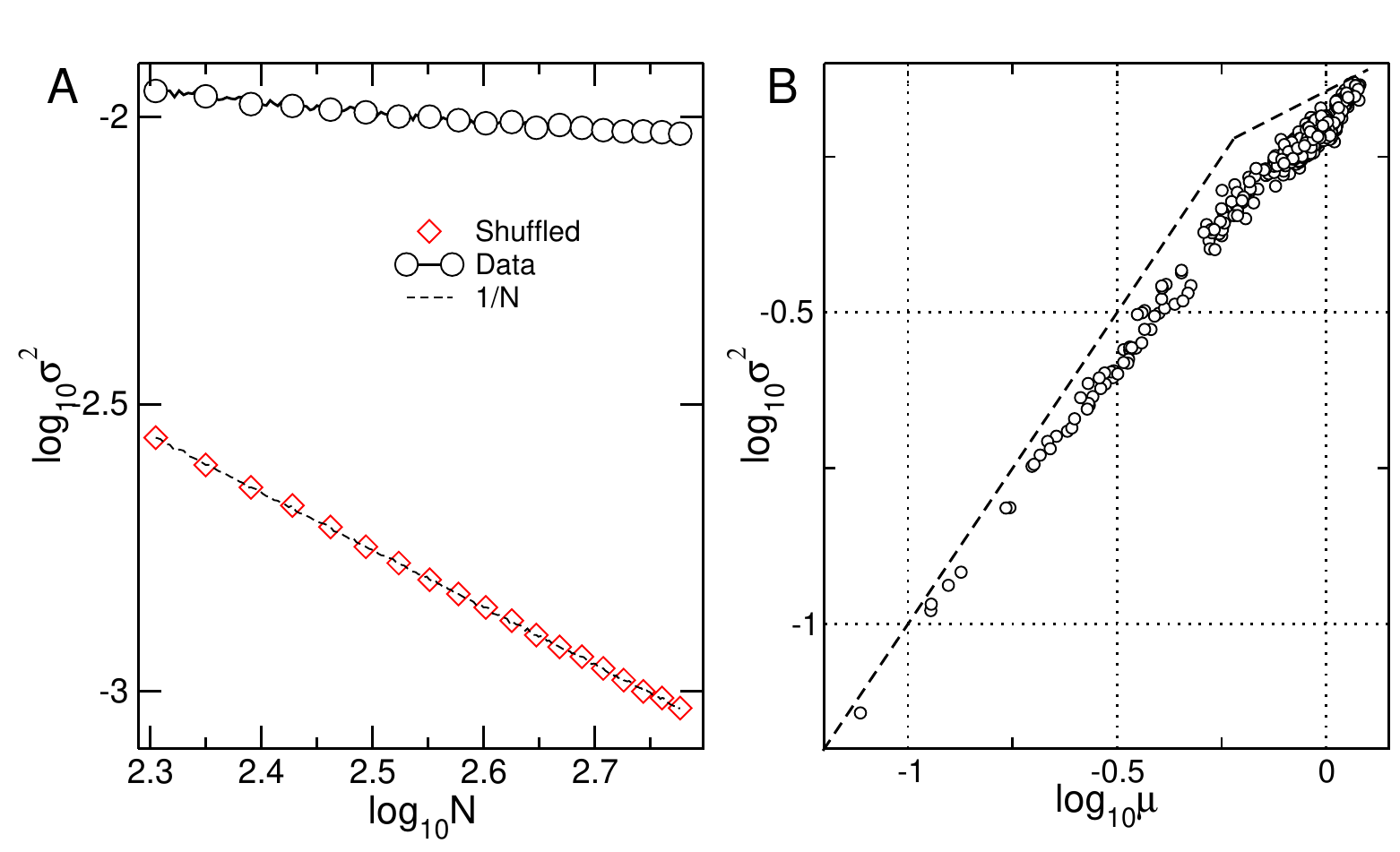}
     \centering
\caption{Fluctuation scaling of the bees density reflects the presence of correlations. Panel A:  Finite-size scaling of the variance as a function of the ensemble size $N$. Each point corresponds to the variance computed from density fluctuations of ensembles of N grid units (averages of 30 random realizations).
The fluctuation of the raw data (black circles) exhibits a relatively constant variance while the randomized data (red diamonds) follows 1/N scaling as expected from the law of the large numbers.         
Panel B: Taylor Law.  The variance $\sigma^2$ of the density fluctuations in each grid site   as a function of its average $\mu$  does not follow the proportionality of an independent uncorrelated process. The scaling can be approximated by two power-laws with exponents: $\alpha_1\propto 0.89$ and $\alpha_2 \propto 0.59$. Crossover takes place near $\log_{10}\mu=-0.22$. Dashed lines correspond to slopes 1 and 1/2 and serve for visual reference. Both panels represent the data from Phase I.}
\label{fig:Variance}
\end{figure}
Now we turn to explore the collective properties of the fluctuations described. The first indication is given by the fluctuations in the polarization $\Phi$, a parameter commonly used to quantify the degree of global order in collective motion \cite{Midges}. This metric ranges from 0 to 1 and is higher when directions of individual velocity vectors $\vec{v_{i}}$ are aligned. 
\[
\Phi=\bigg| \bigg | \frac{1}{N}\sum_{i=1}^{i=N}\frac{\vec{v_{i}}}{||v_i||} \bigg| \bigg|
\tag{1}
\label{eq:Polar}
\]
By simple visual inspection of the traces in Figure \ref{fig:Variance} it can be noticed that the increases in the nearest neighbor variance are associated with an increase in the polarization (although less obvious it seems that the increases in polarization seem  to precede the spread of the bees nearest neighbor distances). 
Also is also worth mentioning that the polarization and speed seem to hold an inverse relation, an observation that will be commented later on.
These results show that more packed conditions (i.e., smaller NN distances) and less polarized collective motion came together with slower speeds on average. Also during more polarized motion bees spread apart only slightly on the average (observe only a small change in the mean NN distance) but with notable non-uniformity.

Another approach to reveal emergent properties analyzes different statistics computed over increasingly larger ensembles. The results of this type of finite-size scaling analysis can distinguish collective properties when compared with a known null hypothesis in which there are no collective effects. For that, we use the time series of bee densities $x(t)$ recorded at each of the 600 square grids. After constructing ensembles of increasing size  N we computed the statistics of  $f(t)=1/N \sum_{i=1}^N x(t) $ for several different stochastic combinations.
The average results on Figure \ref{fig:Variance}A demonstrate a manifest anomalous scaling of the $f(t)$ variance $\sigma^2$ with increasing $N$, since it remains almost constant. This scaling is in clear contrast with the $1/N$ behavior of an independent hypothesis constructed by random scrambling of the same time series. Together with the polarization calculations already commented on, this result is an indication of collective behavior since it shows that the density fluctuations inside one grid are not independent of the other grids fluctuations. In addition, we explored the so-called fluctuation scaling \cite{menezes}, also known as Taylor law,\cite{eiler}, which considers the proportionality between the mean and the variance of the fluctuations in each grid. The results  shown in Figure \ref{fig:Variance}B indicate that the process can not be due to independent random processes since it shows excess variance with respect to the Poisson hypothesis, demonstrated by the exponent $\alpha > 1/2$ in the  scaling $\sigma^2 \propto \mu ^ {-\alpha} $.

\begin{figure}[t]
    \includegraphics[width=.9\columnwidth]{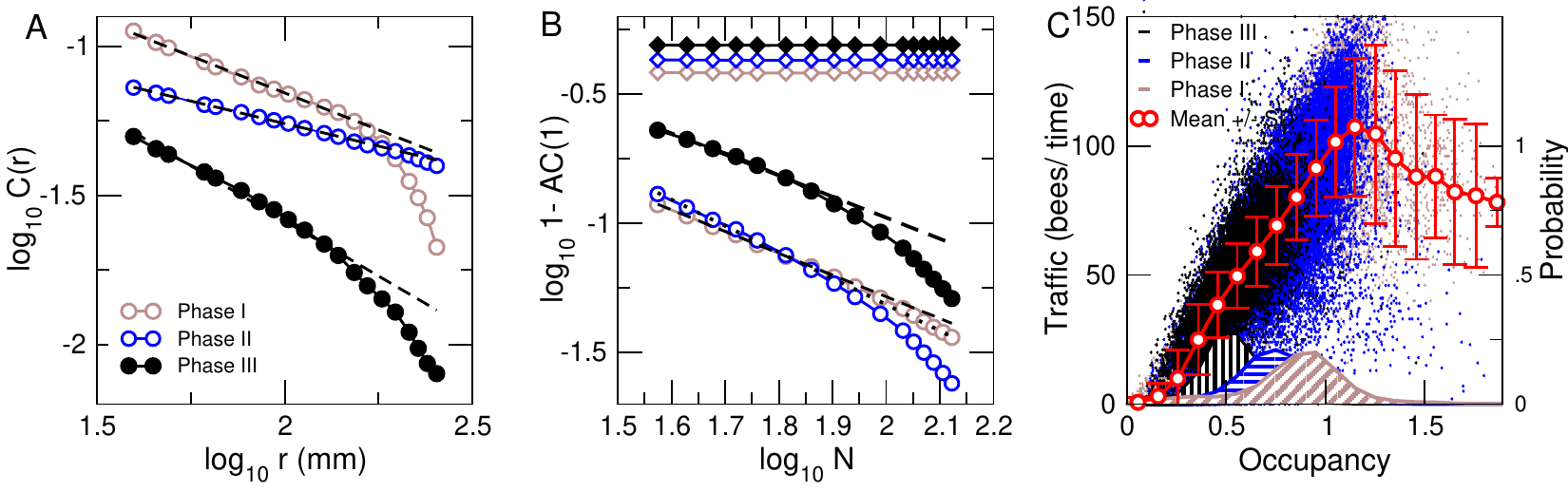}
     \centering
    \caption{Spatial and temporal correlations of the occupancy fluctuations are long-range and dependent on bee density.  
 Panel A: Average  correlation function $C(r)$ of the density fluctuations versus distance $r$ in mm. Dashed lines indicate  fitted slopes, which are equal to -0.5, -0.3, and -0.75 for phases I, II, and III respectively. 
 Panel B: Average  first coefficient of the autocorrelation function $1-AC(1)$ as a function of the size of the ensemble $N$ (number of grid sites) considered. Each point represents the average  of 30 stochastic realizations. Circles represent the results computed from the raw  data, while the diamonds correspond to the computation using the null hypothesis constructed by a circular random shifting of the timeseries. Note that the autocorrelations in the null hypothesis data remain constant irrespective of the ensemble size $N$, while in the raw data AC(1) increases as a power law as the size N of the ensemble increases. Dashed lines represent slopes, which are equal to -0.84, -1.15, and -0.7 for the respective phases. 
Panel C shows the bees' displacement (in analogy to ``traffic'') as a function of density.
The color distinguishes different phases of the experiment:  brown (phase I) before the hive was open, black (phase III), after foragers have been removed from the hive and blue the remaining part (phase II).  The red line with white circles represents the binned average (+/- SD) computed for all the points irrespective of the phase.  The histograms at the bottom of the panel C represent the computed  probability that such density was observed for each phase, providing for an estimation of the residence time of the hive with those conditions. }
    \label{fig:traffic}
\end{figure}

\emph{Density-Density correlations:} Information relevant to untangle collective phenomena  usually can be grasped from the computation of the correlation functions in time \eqref{eq:Temporal_correlations} and in space \eqref{eq:Space_correlations}
\[
AC(k)=\frac{\sum_{t=k+1}^{T}(x_{t}-\langle x\rangle)(x_{t-k}-\langle x \rangle)}{\sum_{t=1}^{T}(x_t-\langle x \rangle)^2} 
\tag{2}
\label{eq:Temporal_correlations}
\]
\[
C(r)=\frac{\sum (x_i(r)-\langle x(r) \rangle)(y_i(r)-\langle y(r) \rangle)}
{\sqrt{\sum (x_i(r)-\langle x(r) \rangle)^2}\sqrt{\sum(y_i(r)-\langle y(r) \rangle)^2}}
\tag{3}
\label{eq:Space_correlations}
\]

 between the occupancy time series. In particular,  the correlation decay as a function of distance, as presented in Figure \ref{fig:traffic}A, where it can be seen that the fluctuations between two points in the hive are correlated an order of magnitude beyond the spatial scale the bees interact ($\sim 8 mm$, see Figure \ref{fig:Fig2}).

 It is expected, as in other complex systems near a critical point, that the properties of the correlations in space reflect into the temporal correlations \cite{grigera,camargo}. Thus, in Figure \ref{fig:traffic}B we estimate the behavior of the autocorrelation function as a function of the ensemble size N. The results show that the temporal correlations of larger ensembles become asymptotically longer, i.e., 1-AC(1) vanishes, in contrast with the null hypothesis results (depicted with diamonds) which remain larger and constant.

In addition, the correlation results in Panels A and B show an important feature: when the density in the hive decreases in phase III (due to the removal of foragers) the correlation properties seem to change to a regime that can be considered of relatively larger independence. Indeed, on the spatial aspects, the correlations shift to a weaker and shorter range (see filled circles in Panel A) which agrees with the temporal aspects where the autocorrelations become also weaker and shorter (see filled circles and diamonds in Panel B). 
 \begin{figure}
    \centering
    \includegraphics[width=.4\columnwidth]{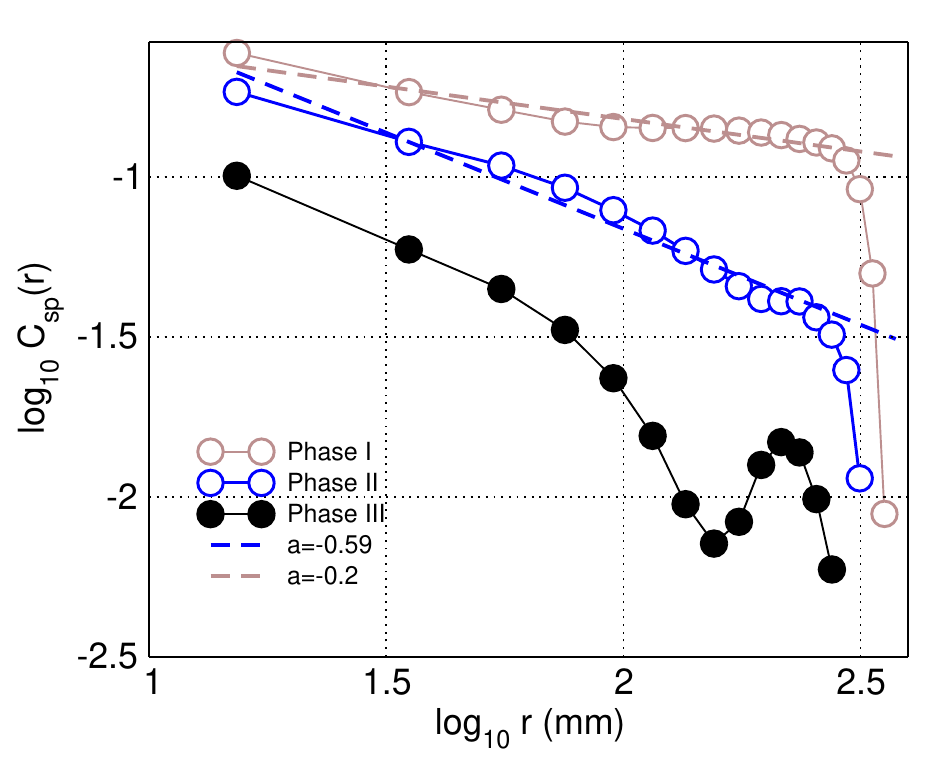}
    \caption{The speed spatial correlations fluctuations are also long-range and dependent on bee density. Speed correlation function: Distance-binned correlations in speed computed using 48-hour periods at the three phases of the experiment. (30 mm bins are used, and median values of the bins are used to represent distance on x-axis)} 
   \label{fig:FigCorr}
\end{figure}
We conjecture that the presence of density-dependent correlations may shed light on the mechanism responsible for the collective dynamics. To test that we computed an observable alike to traffic by first partitioning the timeseries into 5000-sec intervals and computing the average occupancy of each grid square. For the same intervals, we compute the throughput of bees by dividing each square with a midline and counting the occurrences of bees crossing this line in either direction.  The results, presented in Fig.\ref{fig:traffic}, unveil a pattern that is interpreted as follows: For low values, the traffic is a linear function of occupancy. This is true for all phases when the occupancy within the interval is lower, but it is especially evident if only the data collected in phase III (black dots) is considered because there is one fold reduction in density due to the foragers removal (see also Fig.\ref{fig:Fig2} ). For higher densities, there is a maximum in the traffic, beyond which any further increase in density leads to a reduction in the bee's traffic. Considering the typical size of a bee  ($\sim 10$ mm) and the mean distance to the nearest neighbor computed ($\sim 7-9 $ mm, see Fig\ref{fig:Fig2}A) the observed maximum traffic occurs near a regime in which  bees would only be able to move as a solid-like ensemble. 

The changes in the traffic and the correlation properties seen in Fig.\ref{fig:traffic}, together with the  changes in the nearest neighbor distances shown in Fig.\ref{fig:Fig2}A  seem to be interrelated. As a first hypothesis, these results may be explained by  a process of jamming-unjamming as described in a variety of systems from car traffic \cite{traffic} to cells \cite{jammingcells}. Jamming is a self-organized process found both in living and inert systems in which even a  small change in the control parameter, such as density, can induce big changes in the macroscopic properties of the system or material. 


\emph {Speed correlations:} Additional evidence is provided by the results in Figure \ref{fig:FigCorr}, which presents the correlations in speed, computed for all bees present in the hive during different phases of the experiment. Speed is computed as $v=||\frac{\delta \vec{r}}{\delta t}||$; $\delta t =5 \; sec$. Distance-binned speed correlation coefficient $C_{sp}(r_1, r_2)$ for each bin is computed as a mean correlation coefficient  between the speed time series $v_i, v_{j}$ of all bee pairs located at mutual distance $r$: $r_1<r<r_2$, were $r_1$ and $r_2$ are the bin' edges \eqref{eq:Speed_cor}. 
\[
\tag{4}
\label{eq:Speed_cor}
C_{sp}(r_1, r_2)=\frac{1}{\sum_{i\neq j}\theta(||\vec{r_{i}}-\vec{r_{j}}||)}\sum_{i\neq j}\frac{\sum_{1}^{t}(v^t_i-\langle v_i\rangle)(v^t_j-\langle v_j\rangle)\theta(||\vec{r^t_i}-\vec{r^t_j}||)}{\sqrt{\sum_{1}^{t} (v^t_i-\langle v_i \rangle)^2}\sqrt{\sum_{1}^{t}(v^t_j-\langle v_j \rangle)^2}} \\
\]
\[
\theta=
        \begin{cases}
        1 & \text{if } r_1<||\vec{r_i}-\vec{r_j}||<r_2 \\ 
        0 & \text{otherwise } \\ 
        \end{cases} 
\]

Results demonstrate that the spatial correlations in \emph{speed} are only long-range (i.e., decaying slowly with distance as a power law) for when the bee density in the hive is high enough (Phases I and II) and become short-range (i.e., decaying faster) when the density of bees is reduced in Phase III such that their interactions are weaker.   
  
 In summary, we have shown that bees in the hive exhibit complex collective dynamics that involve scale-free spatial and temporal correlations. We conjecture possible mechanisms, through which such dynamics could emerge. Several routes are open for further study including appropriate modeling efforts concerning the dynamics of jamming of active particles in 2D. Jamming transitions are well investigated with respect to traffic phenomena, however one would expect additional effects coming from behavior germane to eusocial insects, as suggested previously \cite{Ants_Move}.   
 
 The correlation properties observed for the fluctuations in the hive might indicate that the system is tuning itself towards an optimal state by varying its control parameter - bee density. This critical state corresponds to a density that provides maximum throughput (i.e., peak at Fig.\ref{fig:traffic}). It is striking that such throughput optimization is very similar to what car drivers experience in daily traffic, including the presence of long-range correlations in speed and occupancy, much larger than the the short-range interactions. The present results call for modeling attempts to validate this conjecture.

 \emph{Acknowledgments:}  We are indebted to the authors of Ref.\cite{Gernart} for sharing the data and to T. Gernart for help in clarifying data and recording details. I.S. thanks the hospitality of UNSAM and to the members of the Instituto de Ciencias F\'isicas, where this work was conducted. I.S. was supported by funding from ECSU Unit in OIST. This research was supported by Grant No. 1U19NS107464-01 from NIH BRAIN Initiative, by CONICET (Argentina) and Escuela de Ciencia y Tecnolog\'ia, UNSAM.



\begin{thebibliography}{99}


\bibitem{Wilson1} E. O. Wilson, Chemical communication among workers of the fire ant Solenopsis saevissima (Fr. Smith) 1. The Organization of Mass-Foraging,  {\em Animal Behavior} 10, 134-164 (1962). 

\bibitem{Wilson2}E. O. Wilson, The Insect Societies, Cambridge: Belknam Press, 1971.

\bibitem{Wilson3}E. O. Wilson, Sociobiology, Cambridge: Belknam Press, 1975.


\bibitem{rauch} E. Rauch, M. M. Millonas, D.R. Chialvo, Pattern formation and functionality in swarm models, {\em Physics Letters} A 207: 185-193 (1995).

 \bibitem{Cavagna1}A. Cavagna, A. Cimarelli, I. Giardina, G. Parisi, R. Santagati, F. Stefanini, M. Viale, Scale-free correlations in starling flocks, {\em Proc. Natl. Acad. Sci.} 107, 11865 (2010).

\bibitem{Cavagna2} A. Attanasi, A. Cavagna, L. Del Castello, I. Giardina, S. Melillo, L. Parisi, O. Pohl, B. Rossaro, E. Shen, E. Silvestri,  M. Viale, Finite-Size Scaling as a Way to Probe Near-Criticality in Natural Swarms, {\em Phys. Rev. Lett.} 113, 238102 (2014).
\bibitem{shimmering} S. Vijayan,  E. J. Warrant, H. Somanathan. Defensive shimmering responses in Apis dorsata are triggered by dark stimuli moving against a bright background. {\em Journal of Experimental Biology.} 225 (17)  jeb244716. (2022).



\bibitem{Cavagna3}A. Cavagna, I. Giardina, T. S. Grigera, The physics of flocking: Correlation as a compass from experiments to theory, {\em Physics Reports} 728, 1 (2018).

 \bibitem{peleg}O. Shishkov and O. Peleg, Social insects and beyond: The physics of soft, dense invertebrate aggregations. {\em Collective Intelligence} 1 (2), 1-18 (2022).



\bibitem{Gernart} T. Gernat, V.D. Rao, M. Middendorf, H. Dankowicz, N. Goldenfeld, G.E. Robinson, . Automated monitoring of behavior reveals bursty interaction patterns and rapid spreading dynamics in honeybee social networks.  {\em Proc. Natl. Acad. Sci.}, 115(7), 1433-1438 (2018).

\bibitem{chialvo1}D.R. Chialvo,  Emergent complex neural dynamics  {\em Nature Physics} 
   6,  744-750 (2010).

\bibitem{menezes} M.A. de Menezes and A.-L. Barabasi, Fluctuations in network dynamics. {\em Phys. Rev. Lett.}92, 28701 (2004).

\bibitem{eiler} Z. Eisler and J. Kertesz,  Scaling theory of temporal correlations and size-dependent fluctuations in the traded value of stocks {\em Phys. Rev. E} 73, 046109 (2006).

\bibitem{grigera}T. S.  Grigera, Correlation functions as a tool to study collective behaviour phenomena in biological systems. {\em Journal of Physics: Complexity}, 2(4), 045016 (2021).

\bibitem{camargo} S. Camargo,  D.A. Martin, E.J.A. Trejo, A. de Florian, M.A. Nowak, S.A. Cannas, T.S. Grigera, D.R.  Chialvo, Scale-free correlations in the dynamics of a small ($N\sim10000$) cortical network. {\em Physical Review E}, 108(3), 034302 (2023).

\bibitem{traffic} K. Nagel and  M. Schreckenberg,   A cellular automaton model for freeway traffic. {\em Journal de Physique I} 2 (12) 2221 (1992).


\bibitem{jammingcells} O. Chepizhko, C. Giampietro, E. Mastrapasqua, M. Nourazar, M. Ascagni, M. Sugni, U. Fascio, L. Leggio, C. Malinverno, G. Scita, S. Santucci, M. J. Alava, S. Zapperi, C.A. M. La Porta, Bursts of activity in collective cell migration, {\em Proc. Natl. Acad. Sci.} 113(41) 11408-11413 (2016).

\bibitem{Ants_Move} R. Gallotti and D.R. Chialvo, How ants move: individual and collective scaling properties. \emph{Journal of The Royal Society Interface}, 15(143), 20180223 (2018).

\bibitem{Fine_Tuning} D.R. Chialvo, S.A. Cannas, T.S. Grigera, D.A. Martin, D. Plenz,  Controlling a complex system near its critical point via temporal correlations. \emph{Scientific reports}, 10(1), 12145 (2020).
\bibitem{Midges} A. Attanasi, A. Cavagna, L. Del Castello, I. Giardina, S. Melillo, L. Parisi, O. Pohl, B. Rossaro, E. Shen, E. Silvestri, M. Viale,  Collective behaviour without collective order in wild swarms of midges. \emph{PLoS computational biology}, 10(7), e1003697 (2014).
\bibitem{Honeybees_book} T.D. Seeley,  Honeybee democracy. \emph{Princeton University Press}, 2011.
 \end{thebibliography}
\end{document}